\documentclass[apj]{emulateapj}
\usepackage{graphicx, graphics}
\usepackage{natbib}
\usepackage{amsmath,amssymb}

\received{}
\revised{}
\accepted{}

\newcommand{\chandra}{{\sl Chandra}~}
\newcommand{\wise}{{\sl WISE}~}
\newcommand{\swift}{{\sl Swift}~}
\newcommand{\galex}{{\sl GALEX}}
\newcommand{\xmm}{{\sl XMM}-Newton~}

\begin{document}
\title{On the Host Galaxy of GRB 150101B and the Associated Active Galactic Nucleus}
\author{Chen Xie\altaffilmark{1}, Taotao Fang\altaffilmark{2,3}, Junfeng Wang\altaffilmark{2,3}, Tong Liu\altaffilmark{2,3,4}, and Xiaochuan Jiang\altaffilmark{2}}
\altaffiltext{1}{Department of Physics, Xiamen University}
\altaffiltext{2}{Department of Astronomy and Institute of Theoretical Physics and Astrophysics, Xiamen University; fangt@xmu.edu.cn}
\altaffiltext{3}{SHAO-XMU Joint Center for Astrophysics}
\altaffiltext{4}{Department of Physics and Astronomy, University of Nevada, Las Vegas, NV 89154, USA}
\begin{abstract}

We present a multi-wavelength analysis of the host galaxy of short-duration gamma-ray burst (GRB) 150101B. Follow-up optical and X-ray observations suggested that the host galaxy, 2MASX~J12320498-1056010, likely harbors a low-luminosity active galactic nuclei (AGN). Our modeling of the spectral energy distribution (SED) has confirmed the nature of the AGN, making it the first reported GRB host that contains an AGN.  We have also found the host galaxy is a massive elliptical galaxy with stellar population of $\sim 5.7\ Gyr$, one of the oldest among the short-duration GRB hosts. Our analysis suggests that the host galaxy can be classified as an X-ray bright, optically normal galaxy (XBONG), and the central AGN is likely dominated by a radiatively inefficient accretion flow (RIAF). Our work explores interesting connection that may exist between GRB and AGN activities of the host galaxy, which can help understand the host environment of the GRB events and the roles of AGN feedback. 

\end{abstract}
                                                                                                                                                         

\section{Introduction}

Gamma-ray bursts (GRBs) are transient events that can be divided into two classes, short-duration ($\textless$ 2 s) and long-duration ($\textgreater$ 2 s) \citep{Kouveliotou et al.(1993)}, or Type I and II GRBs \citep{Zhang(2006), Zhang et al.(2009)}. The physical origins of the two classes are different: long GRBs (LGRBs) are trigged by the collapse of low metallicity, massive star \citep{Woosley(1993), Woosley  Bloom(2006) 2006ARA&A}, and short GRBs (SGRBs) are thought to be originated from mergers of compact binary systems, such as a black hole and a neutron star (BH-NS), or two neutron stars (NS-NS) \citep{Eichler et al.(1989), Paczynski(1991), Narayan et al.(1992)}. 

The environments of the GRB host galaxies provide important clues for understanding these energetic cosmic events, and have been under intensive study in the past two decades (see, e.g. \citealp{Gehrels et al.(2009)} and \citealp{Berger(2014)} for reviews). In general, long-duration bursts are associated with star-forming galaxies \citep{Le Floc'h et al.(2006), Chary et al.(2007), Fynbo et al.(2008), Savaglio et al.(2009)} that can be either faint or bright \citep{Perley et al.(2016a), Perley et al.(2016b)}. On the other hand, the short-duration bursts typically have long-lived progenitors, and their host galaxies show a mixed population of early and late-type galaxies, with a wide span of star formation rates \citep{Gehrels et al.(2009), Berger(2014)}.

Up until GRB~150101B, neither long or short-duration GRB hosts show activities of active galactic nuclei (AGNs) in the center. Due to the link between the long-duration bursts and the core-collapse of massive stars, it was suspected that some GRB hosts with bright sub-millimeter or infrared emission may be powered by AGNs (see, e.g., \citealp{Tanvir et al.(2004), Gehrels et al.(2009), Stanway et al.(2015)}). Several attempts have been made to identify AGNs in the GRB hosts but yield no results \citep{2005cxo..prop.2013F, 2011cxo..prop.3720P, 2012xmm..prop..140S}. Yet, it remains extremely interesting to explore the connection between GRBs and the star-formation and AGN activities of the host galaxies, as it may provide important clues about the roles of AGN feedback, as well as the GRB host environment.
 
In this Letter, we present a multi-wavelength analysis of the host galaxy of GRB 150101B. We confirm this host galaxy contains a central AGN, which was first suggested by \citealp{GCN17281}, \citealp{GCN17285}, and \citealp{GCN17289}. To date, it is the first confirmed AGN in known GRB host population. We also suggest that the host galaxy can be classified as an X-ray bright, optically normal galaxy (XBONG, \citealp{Moran et al.(1996)}; \citealp{Yuan Narayan(2004)}; \citealp{Brandt Hasinger(2005)}). Observations and data analysis are presented in Section~2.  In Section~3, we fit the host SED and discuss the nature of the host galaxy and central AGN. Section~4 is discussion and summary. Throughout the paper, we adopt the standard $\Lambda$CDM cosmology with $\Omega_{\rm M}$=0.30, $\Omega_\Lambda$=0.70, and H$_0$=70 km s$^{-1}$ Mpc$^{-1}$.

\section{Multi-wavelength Observations of GRB~150101B and its Host Galaxy}

\subsection{$\gamma$-ray and Optical}

\swift Burst Alert Telescope (BAT) was trigged by GRB 150101B at 15:23 UT on 01 January 2015. The light-curve shows a single peak with a duration (T$_{90}$) of 0.018 $\pm$ 0.006 seconds \citep{GCN17267}.  {\sl Fermi} was also triggered with a duration (T$_{90}$) of about 0.08 s (50-300 keV) \citep{GCN17276}. At $\sim3\arcsec$ away, the Very Large Telescope (VLT) clearly detected the optical emission from the bright galaxy, 2MASX J12320498-1056010,  which was identified as the host galaxy of GRB~150101B \citep{GCN17281}.  The VLT spectra show several prominent absorption features, most notably Mg b and Na D, at a redshift of $0.134$ \citep{GCN17281}. The Gran Telescopio CANARIAS (GTC) also detected a faint, single emission line which they tentatively identified as an [\ion{O}{2}] emission line at $z\sim0.093$ \citep{GCN17278}. However, the GTC data suffered from poor weather condition, passing clouds, and bad seeing ($>$ 2.5$\arcsec$) \citep{GCN17278}. Therefore, we adopt z=0.134 as the redshift of the host galaxy in this work. This is one of the lowest redshifts among SGRB hosts \citep{GCN17333}. The estimated GRB isotropic energy between 15 and 150 keV is 9.7 $\times$ 10$^{47}$ erg at z = 0.134. 

\subsection{X-ray observations}

Two \chandra follow-up observations were performed at 7.8 days \citep{GCN17289} and 39 days \citep{GCN17431} after the burst, respectively. In Figure~1, the images taken by the two \chandra observations show a bright and a fading source. The position of the fading X-ray counterpart is consistent with the only fading optical source in the BAT position, reported by \cite{GCN17333}. 

The X-ray data reduction was performed with  the Chandra Interactive Analysis of Observations (CIAO) software package\footnote{http://asc.harvard.edu/ciao/} (version 4.6). The spectra extracting region is shown in Figure~1 in blue circles. For the X-ray fading counterpart, the net count rates of the first and second epoch are (9.6 $\pm$ 0.8) and (1.4 $\pm$ 0.3) $\times$ 10$^{-3}$ cts/s, respectively. Therefore, the fading X-ray source is identified as the GRB afterglow. 

The position of the bright source is located in the nucleus of the known galaxy 2MASX J12320498-1056010, which was reported by several surveys and follow-up observations (see, Table~1 for photometry).  Adopting the images from the two Chandra observations, the bright source coordinate is $(RA, DEC) = (12^h32^m04.965^s, -10^d56^m00.59^s)$, and the fading source coordinate is $(RA, DEC) = (12^h32^m05.099^s, -10^d56^m02.55^s)$,  with a 90\% uncertainty of 0.6\arcsec. The offset between two sources is $\sim$ 2.77$\arcsec$, and the projected physical offset is 6.61 kpc at z = 0.134. About 35\% SGRBs are located at more than 7 kpc away from the center of the host galaxies (see, Figure~10 in \citealp{Berger(2014)}). Furthermore, the host light extends beyond the position of the GRB optical afterglow \citep{GCN17281}. The possibility of such bright source coincidently appeared in the region of GRB in a radius of 10\arcsec ~is less than 2.4 $\times$ 10$^{-4}$ \citep{Manners et al.(2003)}. Therefore, we conclude that the bright X-ray source is the nucleus of the host galaxy 2MASX J12320498-1056010.

To increase the photon statistics we merged the two observations of the bright source and extract its X-ray spectrum for modeling (green regions in Figure~1). We used the XSPEC package v12.8.2 and adopted a model of power law with two photoelectric absorptions. One photoelectric absorption was fixed at the Galactic value\footnote{http://asc.harvard.edu/toolkit/colden.jsp} of $N(H)=3.24\times10^{20}\rm\ cm^{-2}$, and one was let free at the redshift of the host galaxy. This simple model provides adequate fit to the X-ray spectrum. The fitted $N(H)$ at the redshift of the host galaxy is $3\times10^{19}\rm\ cm^{-2}$, and the power law index is $2.2\pm0.1$. We have derived a flux of 2.86 $\times$ 10$^{-13}$ erg cm$^{-2}$ s$^{-1}$  in 0.5 - 8.0 keV, and a luminosity of 1.37 $\times$ 10$^{43}$ erg s$^{-1}$ in 0.5 - 8.0 keV. As we will discuss later in this paper, the source X-ray luminosity indicates it is a low luminosity AGN, and this AGN showed no variation during the two epochs (32 days). Our flux is somewhat lower than that of \citet{GCN17289}, which may be attributed to the size of the source extraction region.

Note that the \xmm also have $\sim$ 30 ks observation for this source after burst proposed by \cite{GCN17318}. But the GRB and the central AGN cannot be spatial resolved.

\begin{figure}
\center
\includegraphics[height=0.173\textheight,width=0.45\textwidth]{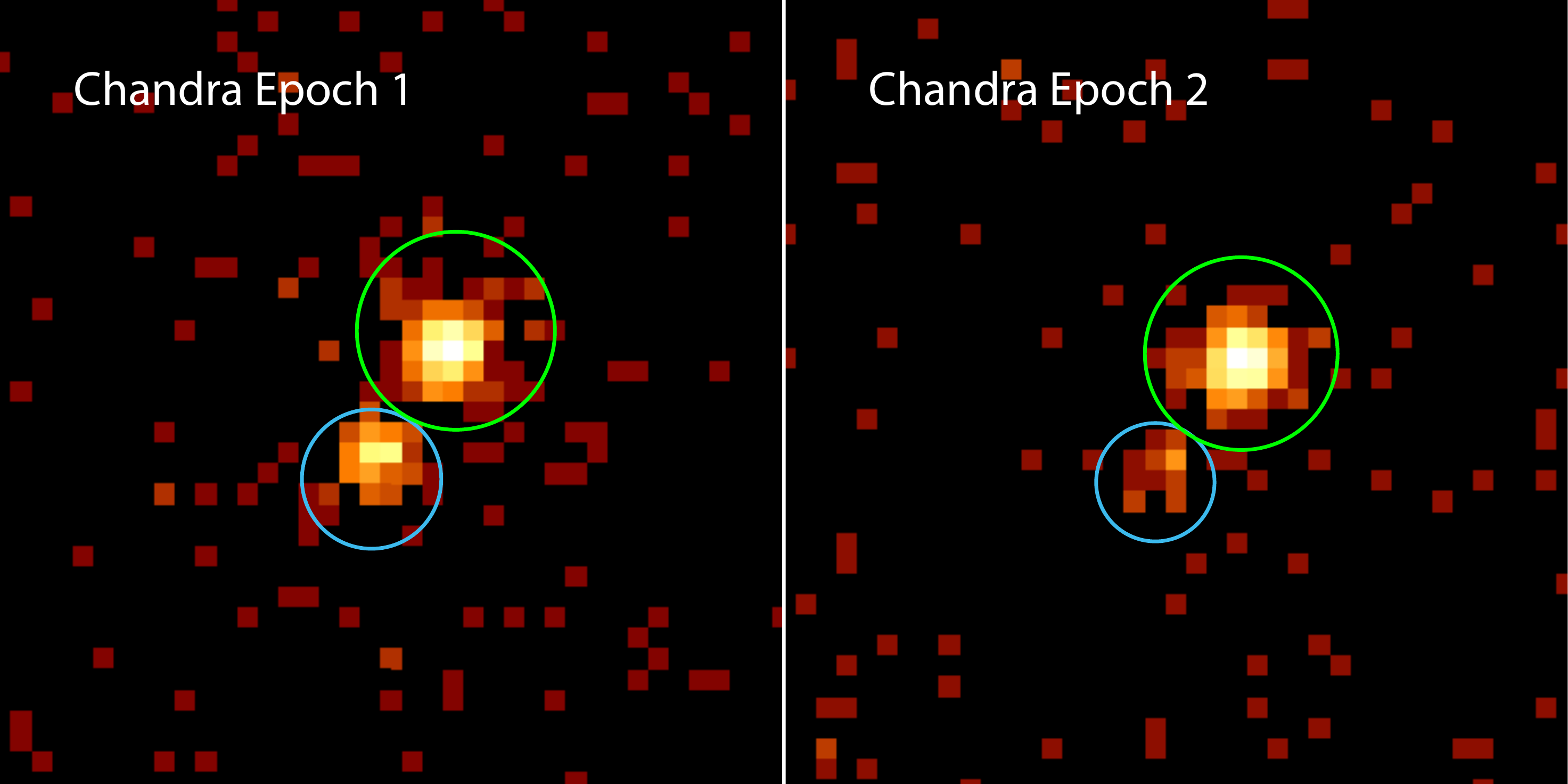}
\caption{Two \chandra observations. The first epoch \citep{GCN17289} began on 09 Jan 2015 (7.83 days post-burst) for an exposure time of 14.87 ks. The second epoch  \citep{GCN17431} began on 10 Feb 2015 (39 days post-burst and 32 days after the first epoch) for an exposure time of 14.86 ks. Comparing the two epochs, we can clearly see the fading afterglow counterpart. AGN (bright and marked by green circles) and afterglow (faint and marked by blue circles) positions are consistent with the host nucleus and GRB optical afterglow, respectively. The circled areas are the regions we extract the spectra of AGN and GRB.
 } 
\label{f1}
\end{figure}

\subsection{Other observations}

Table~1 lists multi-wavelength observations of 2MASX J12320498-1056010, the host galaxy of GRB 150101B. Column~1 shows the observing bands from UV to radio. Column 2 shows the effective wavelength of each band. Column 3 shows the instruments with which observation was taken at each band. Column 4 shows the host fluxes after Galactic extinction correction (except radio fluxes) which A$_{\rm v}$=0.1096 \citep{Cardelli et al.(1989), 2011ApJ...737..103S, 2011MNRAS.413.2570R}. The Galaxy Evolution Explorer (\galex) have detected the host, namely GALEX J123205.0-105600, at distance of 0.6\arcsec~ from the center of 2MASX J12320498-1056010. The optical/UV telescope (OM) on board of \xmm and Wide-field Infrared Survey Explorer ({\sl WISE}; \citealp{Wright et al.(2010)}) have also detected the host at distance of 2.7\arcsec~and 0.4\arcsec, respectively. The radio continuum at 9.8 GHz and 4.9 GHz are adopted from \cite{GCN17288, GCN17286}. The 1.4 GHz flux is taken from he NRAO VLA Sky Survey (NVSS; \citealp{Condon et al.(1998)}).

\begin{deluxetable}{lcccccccccc}
\tablewidth{0pt}
\tabletypesize{\scriptsize}
\tiny
\tabletypesize{\footnotesize}
\tablecolumns{30}
\tablecaption{GRB 150101B Host observed photometry
}
\tablehead{
\colhead{Band} &
\colhead{$\lambda_{eff}$} &
\colhead{Instrument} &
\colhead{Flux (mJy)$^{ab}$} &
}
\startdata
FUV& 1528 $\rm{\AA}$ &GALEX & 0.013 $\pm$ 0.005   \\
NUV& 2271 $\rm{\AA}$ &GALEX & 0.012 $\pm$ 0.003   \\
UVM2 & 2310 $\rm{\AA}$ & XMM\ OM & $\textless$ 0.013$^{c}$  \\
UVW1 & 2910 $\rm{\AA}$ & XMM\ OM & $\textless$ 0.022$^{c}$\\
U & 3440 $\rm{\AA}$ & XMM\ OM & $\textless$  0.046$^{c}$ \\
V & 5430 $\rm{\AA}$ & XMM\ OM & $\textless$ 0.636$^{c}$ \\
J& 1.24 $\mu$m &2MASS& 1.120 $\pm$ 0.090  \\
H& 1.66 $\mu$m &2MASS& 1.444 $\pm$ 0.144 \\
Ks& 2.16 $\mu$m &2MASS& 1.603 $\pm$ 0.130  \\
W1& 3.4 $\mu$m &WISE& 1.306 $\pm$ 0.032  \\
W2& 4.6 $\mu$m &WISE& 0.917 $\pm$ 0.025  \\
W3& 12  $\mu$m &WISE& 0.323 $\pm$ 0.145  \\
W4& 22 $\mu$m &WISE&  $\textless$ 2.553$^{d}$  \\
9.8 GHz& 3 cm &VLA&  3.15 $\pm$ 0.02  \\
4.9 GHz& 6 cm &WSRT& 7.21 $\pm$ 0.07  \\
1.4 GHz& 20 cm &VLA& 10.2 $\pm$ 1.0  
\enddata
\tablenotetext{a}{Fluxes have been corrected for Galactic extinction (except radio fluxes) which A$_{\rm v}$=0.1096.}
\tablenotetext{b}{Errors are quoted at 1$\sigma$ level unless otherwise specified.}
\tablenotetext{c}{Data taken from \xmm OM were adopted as upper limits since OM cannot separate GRB and the host.}
\tablenotetext{d}{Correspond to 95\% upper limit.}
\end{deluxetable}

\begin{deluxetable}{lcccccc}     
\tablewidth{0pt}
\tabletypesize{\scriptsize}
\tiny
\tabletypesize{\footnotesize}
\tablecolumns{30}
\tablecaption{
}
\tablehead{
\colhead{Proxy} &
\colhead{SFR/M$_{\odot}$ y r$^{-1}$} &
\colhead{conversion factor} 
}
\startdata
SED& 0.057 $\pm$ 0.003 &   --\\
FUV & 0.86 $\pm 0.35 $ &    \cite{Kennicutt(1998)}\\
NUV & 0.83 $\pm 0.20 $ &     \cite{Kennicutt(1998)}\\
12 $\mu m$ & 0.49 $\pm 0.22 $ &  \cite{Jarrett et al.(2013)}\\
22 $\mu m$ & $\textless$ 3.23 &  \cite{Jarrett et al.(2013)}

\enddata
\tablenotetext{*}{Note that the SFRs already been corrected by Galactic extinction \citep{Cardelli et al.(1989), 2011ApJ...737..103S, 2011MNRAS.413.2570R}.}
\end{deluxetable}

\section{The Host  galaxy of GRB 150101B}

\subsection{SED Fitting and the Host Galaxy}

Using all the available photometric data presented in Table~1, we have fitted the spectral energy distributions (SEDs) of the host galaxy with the Code Investigating GALaxy Emission (CIGALE; \citealp{Noll et al.(2009)}) \footnote{http://cigale.lam.fr/about/}. We have also included an AGN component \citep{Fritz et al.(2006)}. In Figure~2 we present the data points and the model continuum. The fit is reasonably good with a $\chi^2/d.o.f$ of $9.6/11$.

In Table~2 we list the star formation rates of the host galaxy, estimated using SED, UV and infrared data. Both UV and IR data show star formation rates that are significantly higher than those estimated by the SED fitting, suggesting strong contamination by the central AGN.

Fitting of the SED of the host galaxy suggests a stellar mass of log$_{10}$(M$_{\star}$/M$_{\odot}$) =  11.07 $\pm$ 0.06, and a galaxy age of 5.7 $\pm$ 1.0 Gyr. This makes the host galaxy of GRB 150101B one of the oldest SGRB hosts detected so far \citep{Berger(2014)}. The host Av derived from SED is 0.33 $\pm$ 0.02, consist with the very small $N(H)$ observed in X-ray.

The SED fitting suggests the host galaxy of GRB~150101B is a massive elliptical galaxy. Our \wise color-color diagram also suggests the host is likely an elliptical galaxy. We have found [w1] - [w2] = 0.25 $\pm$ 0.04 (mag) and [w2] - [w3] = 0.79 $\pm$ 0.49 (mag), falling into the region of elliptical galaxies suggested by the \wise color-color diagram \citep{Wright et al.(2010)}.

\begin{figure}[t]
\center
\includegraphics[height=0.32\textheight,width=0.52\textwidth]{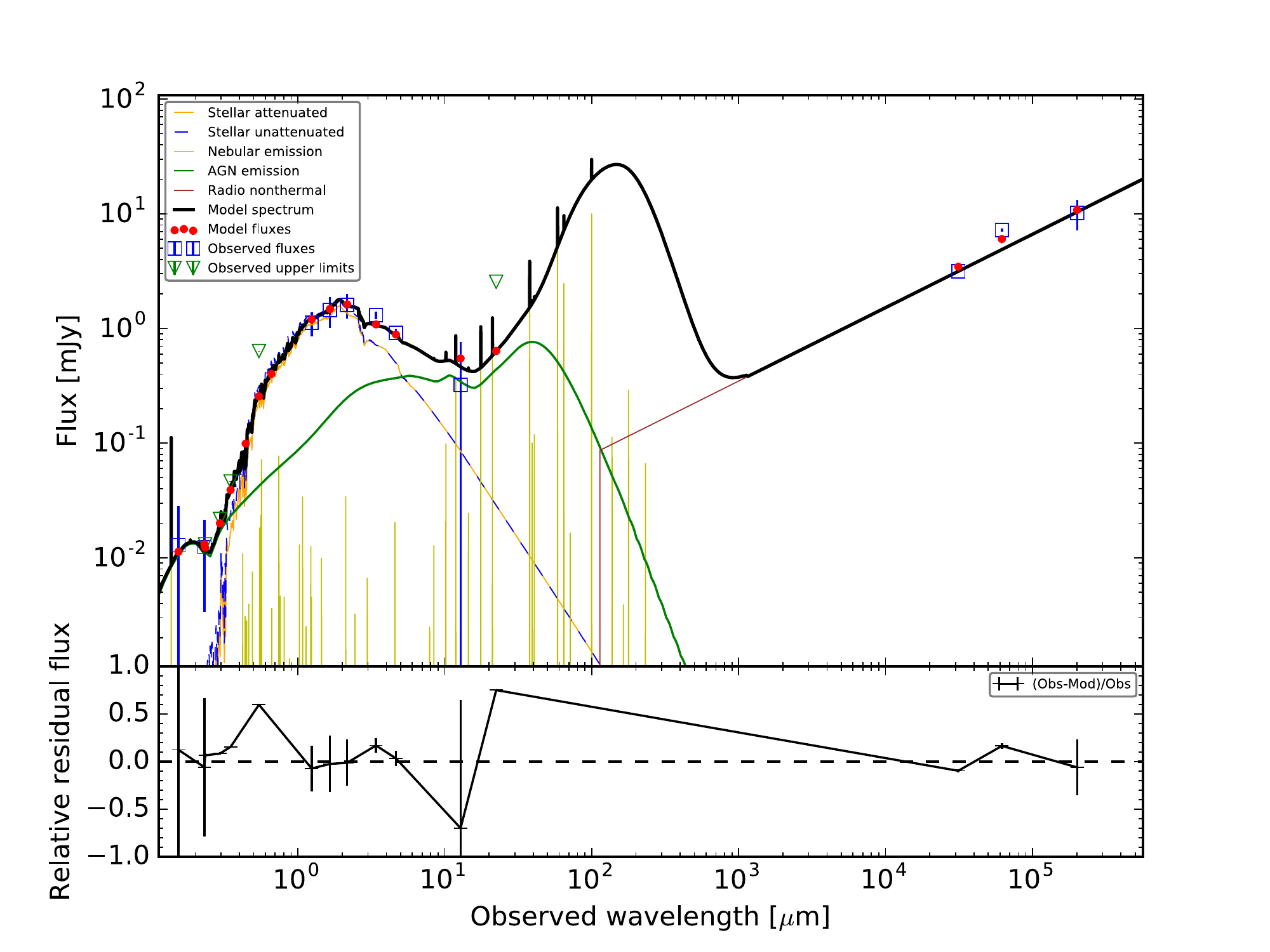}
\caption{SED modeling using CIGALE. The reduced $\chi^2$ is 0.9 with 11 degree of freedom.}
\label{f2}
\end{figure}

\subsection{AGN}

The most unique property is the detection of a central AGN in the host galaxy of GRB~150101B, first suggested by \citealp{GCN17281}, \citealp{GCN17285}, and \citealp{GCN17289}. Our multi-waveband analysis has confirmed the existence of the AGN, which makes it the first GRB host with reported AGN. Beside the clear detection of an X-ray point source in the center of the host galaxy, we draw the conclusion based on the following findings. First, our SED fitting required an AGN component (green line in Figure~2). We have experimented the fit with a model without the AGN component, and have found such fit officially unacceptable ($\chi^2/d.o.f$ = 1.54). The main problem with such model is that it cannot fit the high flux presented in the UV band observed with \galex.

Secondly, the observed X-ray and mid-infrared luminosities are consistent with the intrinsic correlation among AGNs discovered by \cite{Mullaney et al.(2011)}. The observed X-ray luminosity between 2 and 10 keV is $L_X = 7.46\times10^{42}\ {\rm erg\ s^{-1}}$,  and the predicted mid-infrared flux at 12 $\mu m$ is between 0.3 and 4 mJy. This is consistent with the \wise W3 data of 0.323 $\pm$ 0.145 mJy, and suggests that the galaxy emission in mid-IR is dominated by the central AGN. Finally,  the SED modeling also shows a synchrotron emission with a spectral slope of -0.62 $\pm$ 0.03 in the radio band, suggesting AGN-related activities.

Our SED fitting also suggest this is a radio-loud AGN. Conventionally, the radio loudness \citep{Book Netzer(2013)} is defined as: 
\begin{equation}
R=  L_{\upsilon}\textrm{(5~GHz)} / L_{\upsilon}\textrm{(4400~\AA)}.
\end{equation}
Using a B-band flux of 0.099 mJy from our SED model, we have found a radio-loudness of $R \approx 73$. Both the radio loudness and the spectral index in the radio band suggest this AGN is not a type-I AGN. This conclusion is also supported by the derived angle between equatorial axis and line of sight ($48\arcdeg \pm 9 \arcdeg$) from the SED fitting.

Very interestingly, all the properties of the host galaxy of GRB~150101B suggest it belongs to a special class of galaxies, namely the X-ray bright, optically normal galaxies, or XBONGs (see, e.g., \citealp{Moran et al.(1996)}; \citealp{Yuan Narayan(2004)}; \citealp{Brandt Hasinger(2005)}). The observed X-ray luminosity of the host galaxy is consistent with those typical of XBONGs in the range of $10^{41} - 10^{43}\rm\ erg\ s^{-1}$. The observed X-ray to optical flux ratio of $X/O \approx-1.1$ is also consistent those of XBONGs ($X/O \approx-1$). Such flux ratio in general indicates some moderate level of AGN activities. Finally, the lack of optical emission features of those typical of AGNs is also consistent with the definition of XBONGs.

\cite{Brandt Hasinger(2005)} suggested three scenarios that may explain XBONGs: AGNs with heavy obscuration, BL Lac-like objects, or AGNs with radiatively inefficient accretion flow (RIAF, \citealp{Yuan Narayan(2004)}). In the host galaxy of GRB~150101B, the observed low $A_{\rm v}$ and moderate X-ray luminosity suggest the first two scenarios are unlikely. For the central AGN we discussed here,  the observed optical-to-X-ray spectral index is $\alpha_{ox} = 1.15$ (2500$\rm{\AA}$ - 2 keV). Adopting the M$_{\rm BH}$-M$_{\rm Host}$ relation in \cite{Sherman et al.(2014)}, we estimate the M$_{\rm BH} = 10^{8.4} M_{\odot}$ and an Eddington ratio of $\sim 4.4 \times 10^{-3}$. Those values, as well as the radio-loudness, are consistent with the predictions of the RIAF model \citep{Yuan Narayan(2004), Yuan Narayan(2014)}. Therefore, we conclude that the central AGN of the host galaxy is dominated by RIAF.

\section{Discussions and Conclusions} 

In this paper we present multi-wavelength analysis of the host galaxy of GRB 150101B.  We have confirmed the host galaxy harbors an AGN, based on: (1) the X-ray detection of a central point source with high X-ray luminosity; (2) broadband SED modeling; (3) the correlation between observed X-ray and mid-infrared luminosities; and (4) the radio loudness and radio spectral index. We also found the host galaxy belongs to a special class of galaxies, the XBONGs. The observed properties of this AGN are consistent with being dominated by a radiatively inefficient accretion flow. 

GRB 150101B is the first confirmed GRB-AGN system. Up to date no AGN was detected among LGRB hosts, and only one is reported among SGRB hosts (this work). It is still unclear why the GRB host galaxies show very little or no AGN activities. AGNs are rare among galaxies, so one reason might simply be due to counting statistics. The fraction of high-ionization AGN in local galaxies is about 2\% (the fraction of low-ionization AGN is larger) \citep{Book Netzer(2013)}. This rate becomes higher at high redshift when AGNs activities peak between redshift 2 and 3. Up to date, redshift measurements have been reported for a total of $\sim 450$ GRBs\footnote{http://www.astro.caltech.edu/grbox/grbox.php}; however, some of these measurements were performed on the GRB afterglows and not on the host galaxies. Adopting a conservative number of $\sim 200$ GRB host galaxies from the GHostS sample\footnote{http://www.grbhosts.org}, we would expect roughly 4 galaxies with AGN activities if AGN and
GRB are independent phenomena. So it probably is not entirely a surprise if we only observed one. Future multi-wavelength study of the GRB host galaxies should be able to resolve this issue.

On the other hand, if AGNs do become less likely to be detected among the GRB hosts, one possibility is that AGN activities quench the star formation in the host galaxies, therefore reduce the likelihood of detecting LGRB events in such environment. However, since SGRB hosts typically show a wide span of stellar population ages, they clearly exhibit a link with the delayed star formation activities (Berger~2014). Therefore, it is likely to detect some level of AGN activities among SGRB hosts. In particular, we have found the host galaxy of the GRB~150101B belongs to XBONGs, suggesting the direction of further study of the link between AGN activities and GRBs.

\acknowledgments
We thank the anonymous referee for helpful suggestions. We thank Dr. Denis Burgarella for answering our questions about CIGALE. We also thank Drs. Weimin Gu, Mouyuan Sun, Tinggui Wang, Feng Yuan, and Bing Zhang for beneficial discussion. This work was supported by the National Basic Research Program of China (973 Program) under grant 2014CB845800, by the National Natural Science Foundation of China under grants 11233006, 11273021, 11473021, 11473022, 11522323, 11525312 and U1331101, by the Fundamental Research Funds for the Central Universities under grant 20720160023, 20720160024, and 2013121008, and by the Strategic Priority Research Program ``The Emergence of Cosmological Structures" of the Chinese Academy of Sciences, Grant No. XDB09000000.

This research has made use of data obtained from the Chandra Data Archive and the Chandra Source Catalog, and software provided by the Chandra X-ray Center (CXC) in the application packages CIAO.  This publication makes use of data products from the Two Micron All Sky Survey \citep{Skrutskie et al.(2006)}, which is a joint project of the University of Massachusetts and the Infrared Processing and Analysis Center/California Institute of Technology, funded by the National Aeronautics and Space Administration and the National Science Foundation. Based in part on public data from GALEX GR6/GR7. The Galaxy Evolution Explorer (GALEX) satellite is a NASA mission led by the California Institute of Technology. This publication makes use of data products from the Wide-field Infrared Survey Explorer \citep{Wright et al.(2010)}, which is a joint project of the University of California, Los Angeles, and the Jet Propulsion Laboratory/California Institute of Technology, funded by the National Aeronautics and Space Administration. This research has made use of the GHostS database (www.grbhosts.org), which is partly funded by Spitzer/NASA grant RSA Agreement No. 1287913.

\end{document}